\documentclass[%
aapm,
mph,%
amsmath,amssymb,
reprint,%
]{revtex4-2}
\usepackage{graphicx}
\usepackage{dcolumn}
\usepackage{bm}
\usepackage{soul}
\usepackage{siunitx}
\usepackage{xcolor}

\usepackage[mathlines]{lineno}
\modulolinenumbers[5]
\begin{document}
	
	\preprint{AAPM/123-QED}
	
		\title[]{
	 Observation of  laser plasma accelerated electrons with  transverse momentum spread below the thermal level   }

	\author{T.L. Steyn$^1$}
    \email{theunis-lodewyk.steyn@universite-paris-saclay.fr}
	\author{A. Panchal$^2$}
        \author{O. Vasilovici$^1$}
        \author{S. Sch\"obel$^3$}
        \author{P. Ufer$^3$}
        \author{F.M. Herrmann$^3$}
        \author{Y.-Y. Chang$^3$}
        \author{I. Moulanier$^1$}
         \author{M. Masckala$^1$}
          \author{O. Khomyshyn$^1$}
        \author{C. Ballage$^1$}
        \author{M. LaBerge$^3$}
          \author{F. Massimo$^1$}
          \author{S. Dobosz Dufrénoy$^2$}
          \author{U. Schramm$^3$}
          \author{A. Irman$^3$}
          \author{B. Cros$^1$}
          \email{brigitte.cros@cnrs.fr}
	\affiliation{ $^1$ LPGP, CNRS Université Paris Saclay, 91400 Orsay , France  $^2$ Université Paris-Saclay, CEA, LIDYL, 91191 Gif sur Yvette, France }%
    	\affiliation{}
        \affiliation{$^3$ Helmholtz-Zentrum Dresden Rossendorf, Bautzner Landstraße 400, 01328 Dresden, Germany }%

	\date{\today}
	
	\begin{abstract}

Achieving high-quality electron beams from laser-plasma accelerators critically relies on density tailoring to control electron dynamics during injection, acceleration, and extraction. We report on the experimental observation of electron beam acceleration and 3D shaping, in transverse momentum and longitudinal phase space, controlled by plasma density tailoring in a gas cell. It leads to relativistic electron beams with a transverse momentum spread of 0.2 $m_e c$, below the thermal level imprinted  by the process of ionization injection in a laser-driven wakefield accelerator. These beams have a charge of 40 pC at an energy of 190 MeV with 1.9\% energy spread and an rms divergence of 0.54 mrad, resulting in single-peak spectra with a brightness of up to 8 pC/MeV/mrad. Using numerical simulations, we show that the decrease in transverse momentum spread starts in the downramp and continues in a 10 mm long plasma tail at the exit of the gas cell, where it is accompanied by energy dechirping.

	\end{abstract}
	
	\maketitle

In Laser WakeField Acceleration (LWFA) \cite{Esarey2009,Hooker2013}, a high intensity, short laser pulse (fs level) travels through an underdense plasma,  generating plasma waves. Under the effects of the accelerating electric field sustained by these waves,  trapped electrons can gain energies in excess of hundreds of MeV over a few millimeters. 
Several methods to trap relativistic electrons in these wakefields have been demonstrated\cite{Esarey2009,Hooker2013,downer_diagnostics_2018}. One efficient method to inject plasma electrons in the plasma waves is ionization injection \cite{pak_injection_2010,Pollock2011,Chen2012,Mirziae2015,Thaury2015,Couperus2017,Irman2018,Kurz2021,Kirchen2021,Chang2023,Picksley2024}. In this scheme, a high-Z gas like nitrogen dopes a target composed of a low-Z gas like hydrogen or helium. The plasma generated by the ionization of the low-Z gas and the release of the outermost electrons of the high-Z gas, occurring before the arrival of the peak of the laser pulse, provide the medium where laser-driven plasma waves are excited. When the laser intensity exceeds the tunnel ionization threshold of the inner atomic energy levels, their ionization near the peak of the laser pulse places some of the released electrons in a phase of the plasma wave where they can be trapped. 
For selected input laser parameters, this injection mechanism can be triggered and stopped by controlling the laser field amplitude during its propagation,  through coupling to a tailored plasma density distribution. This controlled injection can create narrow energy bandwidth electron beams\cite{Kirchen2021}. 

While the mechanism of ionization injection can release and trap large amounts of electrons\cite{Couperus2017}, it imprints a transverse momentum spread on the electron beams~\cite{schroeder_thermal_2014}. In order to achieve a high spectral brightness, both the energy spread and divergence of these beams need to be minimized. The divergence of the beam can be minimized either by minimizing the momentum spread of the beam transversely to the main axis of propagation or by accelerating the beam to higher energies. One of the key methods for decreasing the transverse momentum spread is the use of passive plasma lenses. They are based on the use of the transverse focusing forces associated with plasma waves to focus a charged particle beam \cite{chen_grand_1987}. 
Such a lens, driven in the plasma following an LWFA stage, has been observed to reduce the electron beam divergence by a factor 2 using several combinations of devices (gas jet after a cell or a jet)~\cite{Kuschel2016,Thaury2015lens}  or in a shaped profile in a gas jet\cite{Chang2023}. In addition, in the region of plasma where the laser intensity has dropped significantly, the electron beam can drive its own wakefield \cite{Lehe2014}, leading to a focusing of the rear part of the beam. Other works\cite{darcy_tunable_2019,wu_phase_2019} in the beam driven regime have shown that dechirping associated with the longitudinal field decreases the beam energy spread.

In this letter, we report the experimental
observation of electron beams with energies above 100 MeV and  transverse momentum spread below the thermal level imparted by ionization injection. Injection and acceleration occur in a 3 mm long gas cell. After exiting the gas cell, 
the electron beam propagates through a 
plasma down ramp, followed by a long, low-density plasma tail (LPT). This leads to a decrease of the transverse momentum spread below the theoretical thermal minimum for ionization injection. After the LPT, the beam has a narrow energy spread and a transverse momentum spread of 0.2~$m_e c$ inferred from an rms divergence of 0.54 mrad at 190 MeV. Numerical simulations give insight into the process through which these low divergence beams are produced and show that the LPT contributes to reducing the energy spread and transverse momentum spread by acting simultaneously as a plasma lens and a plasma dechirper.

\begin{figure}[ht!]
	\centering
	\includegraphics[width=0.99\columnwidth]{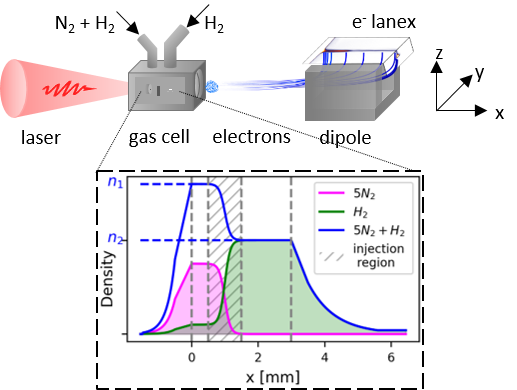} 
		\caption{Schematic of the experimental setup. The laser pulse, propagating along $x$ and linearly polarized along $y$, is focused onto a gas cell target with two compartments, which provides the desired density profile. In the inset, blue: electrons from hydrogen and the first five levels of nitrogen, composing the background plasma; magenta: density of the electrons from the first five levels of nitrogen. $n_1$ and $n_2$ are the plasma electron density on the density plateau of the first and second compartment respectively. The electron beams are characterized with a spectrometer which has a resolution for 150-200~MeV electron beams of 0.045\% and 0.04~mrad for energy and divergence respectively.}
	\label{fig:Exp_Setup}
\end{figure}

This result was obtained using the DRACO laser at Helmholtz-Zentrum Dresden Rossendorf (HZDR), a Ti:Sa laser system with carrier wavelength $\lambda_0=0.8$ $\mu$m providing up to 2.5 J of pulse energy on target, in 30 fs FWHM pulse length with a repetition rate of up to 1 Hz. This laser was focused to a spot size of 24 $\mu$m FWHM into a double compartment gas cell, as illustrated in Fig.~\ref{fig:Exp_Setup}.

\begin{figure}[!ht]
 \centering
 \begin{tabular}{c}
\hspace*{-2mm}\includegraphics[width=1.03\columnwidth]{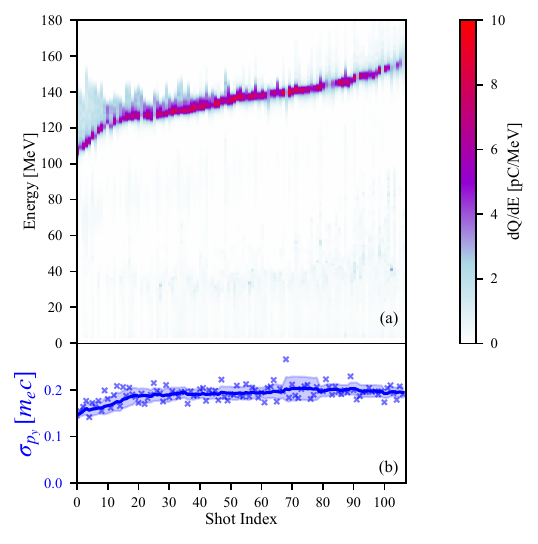}   \\[-10pt]
\hspace*{-2.4mm}\includegraphics[width=1.01\columnwidth]{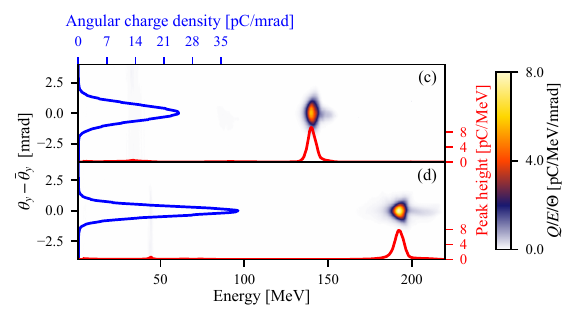}      \\
\end{tabular}
\caption{Experimental results showing the energy spectrum and energy-angle distribution of the measured electron beams. (a) and (b) were obtained with 20 mbar ($n_2 = 10^{18}$ cm$^{-3}$). (a) Energy spectra, ordered by increasing peak energy in the spectrum; (b) normalized transverse momentum spread along $y$ of the same electron beams (blue crosses). The blue line and blue shaded area are the moving average and standard deviation computed across 5 adjacent shots. (c) and (d) are the energy-angle distribution of two electron beams obtained at respectively 20~mbar and 35 mbar ($n_2 = 1.7\times 10^{18}$ cm$^{-3}$) . The pointing ($\bar{\theta}_y$) is  -0.57 and -0.45 mrad respectively. The characteristics of these beams are listed in Table \ref{table:transposed_parameters}. }
\label{fig:Waterfalls}
\end{figure}

We have designed a gas cell with variable and replaceable geometric components, which allows shaping the density profile as illustrated in the inset of Fig.~\ref{fig:Exp_Setup}. A nitrogen-hydrogen mixture is injected into the first compartment  and pure hydrogen into the second compartment. The two filling pressures are controlled independently; pulsed injection of gas at equal pressures ensures proper confinement~\cite{drobniak_two-chamber_2023} of the nitrogen component in the first part of the density profile. The plateau gas densities were obtained from interferometer measurements. The gas density distribution resulting in the plasma density shown in Fig.~\ref{fig:Exp_Setup} was through a parameterization of CFD OpenFOAM \cite{weller1998,openfoam} using the multicomponentFluid package combined with PIMPLE solver allowing the modeling of supersonic multi-component flows.  In addition expected density decay,  a tail  of approximately constant low electron density ($n_{LPT} \approx 4\times 10^{16}$ cm$^{-3}$ for the case of Fig. \ref{fig:Waterfalls} (c)) with length of 10 mm is achieved by shaping the geometry of the outlet.

For  the gas cell shown in Fig.~\ref{fig:Exp_Setup}, 
 optimized electron spectra were achieved at a given focal position ($x = 2.4 $~mm) of the laser beam relative to the gas cell entrance plane. 
 Spectra  having at least 20~pC of charge in the peak are plotted in Fig~\ref{fig:Waterfalls} (a), for a pressure of 20 mbar in both compartments and nitrogen concentration of 15\% in the first compartment, corresponding to plateau densities $n_1 = 1.6 \times 10^{18}$~cm$^{-3}$ and $n_2 = 1 \times 10^{18}$~cm$^{-3}$. These spectra have an average energy of $136 \pm 11$~MeV, and a divergence below 1 mrad. A representative shot in the divergence-energy plane is plotted in Fig~\ref{fig:Waterfalls} (c).  For comparison, a shot at 35 mbar pressure, corresponding to $n_2= 1.7\times 10^{18}$~cm$^{-3}$, is shown in  Fig~\ref{fig:Waterfalls} (d). 

\begin{table*}[!htb]
    \caption{Summary of electron beam parameters. [Exp] and [Sim] refer to measured and simulated parameters extracted at the end of the simulation. $\bar{E}$: energy of the spectrum peak; $\Delta E$: FWHM width of the energy spectrum peak; $Q_T$: total charge in the spectrum; $Q_{FWHM}$: charge in the FWHM peak of the energy spectrum; $\theta_{rms}$: rms divergence along the direction $y$; $\sigma_{p_y}$: rms spread in the normalized transverse momentum along $y$.}
    \label{table:transposed_parameters}
    \renewcommand{\arraystretch}{1.2} 
    \begin{ruledtabular}
    \small
        \begin{tabular}{p{1.cm} p{1.25cm} p{3cm} p{3cm} p{3cm} p{3cm}} 
            &  & 2(a) [Exp] & 2(c) [Exp] & 2(d) [Exp] & 5(b) [Sim] \\
            \hline
            $\bar{E}$ & MeV & $136 \pm 11$ & $140$ & $193$ & $140$ \\
            $\Delta E$ & MeV & $7.1 \pm 2.9$ & $5.6$ & $6.5$ & $9.0$ \\
            $Q_T$ & pC & $64.9 \pm 13.8$ & $70$ & $72$ & $74$ \\
            $Q_{FWHM}$ & pC & $32.0 \pm 7.4$ & $41$ & $41$ & $41$ \\
            $\theta_{\text{rms}}$ & mrad & $0.7 \pm 0.1$ & $0.67$ & $0.46$ & $1.07$ \\
            $\sigma_{p_y}$ & $m_ec$ & $0.19 \pm 0.02$ & $0.18$ & $0.17$ & $0.30$ \\
        \end{tabular}
    \end{ruledtabular}
\end{table*}

For these 107 shots input parameters (laser energy, focus position, gas pressure) were set:  the change in electron energy over the ensemble is due to fluctuations in  plasma density values ($\pm 2.5\%$ in the plateau), and laser parameters (fluctuations of the laser focus are on the order of $\pm 100$ \unit{\micro\meter}). The average and standard deviation of the key characteristics of the beams are given in Table \ref{table:transposed_parameters} column 2(a). These characteristics give peak spectral densities approaching 9 pC/MeV in most cases and spectral brightness peaks approaching 6~pC/MeV/mrad.\\

The  normalized transverse momentum rms spread $\sigma_{p_y}$ plotted  in Fig.~\ref{fig:Waterfalls}(b) was calculated using the measured divergence and energy of the peaked spectra, using
\begin{equation}\label{eq:divergence}
	\sigma_{p_y} = (\beta_y \gamma)_{rms} \approx (\theta_y)_{rms} \bar{\gamma} \, ,
\end{equation}
\vspace{1mm}
under the assumption that the divergence $\theta_y \approx p_y / {p_x} \approx \beta_y$, for ultra-relativistic electrons (i.e. with normalized longitudinal speed $\beta_x\approx 1$), where $p_y$ (respectively $p_x$) is the transverse (longitudinal) momentum. The divergence of electron beams, $(\theta_y)_{rms}$,   is defined as the rms spread of the transverse angle $\theta_y$ computed from the electron distribution within the FWHM of the peak.
$\bar{\gamma}$ is the average of the Lorentz factor $\gamma$ of the electron beam particles. Eq.~\ref{eq:divergence} was obtained assuming a negligible covariance between the Lorentz factor $\gamma$ and the normalized transverse component of the  velocity $\beta_y$ (and between their squares), a negligible average for $\beta_y$, and a variance of the Lorentz factor $Var(\gamma)\ll \bar{\gamma}^2$.

On average, the transverse momentum spread of these beams is $\approx 0.2$ $m_e c$, corresponding to a divergence of $0.72 \pm 0.06  $ \unit{\milli \radian} for the shot ensemble.

For higher operating pressure (see  Fig.~\ref{fig:Waterfalls}(d)), the low transverse momentum spread is maintained as the divergence decreases as the peak energy  (i.e. $\bar{E}$) is increased. In this case, the  density in both compartments was increased by factor of 1.7 (i.e.  $n_2=1.7\times 10^{18}$~cm$^{-3}$ and the focus adjusted to $x= 2.9$ mm, all other parameters were kept unchanged, leading to increased accelerating gradient. The total charge and the charge in the FWHM are very similar to the lower density case, as shown in Table~\ref{table:transposed_parameters}. The transverse momentum spread is 0.18  $m_e c$. 
The decrease of divergence, down to 0.46 mrad, with similar charge and energy spread leads to peaks with an increased spectral brightness of 8 pC/MeV/mrad compared to the 20~mbar case.

\begin{figure}[h]
		\includegraphics[width=0.99\columnwidth]{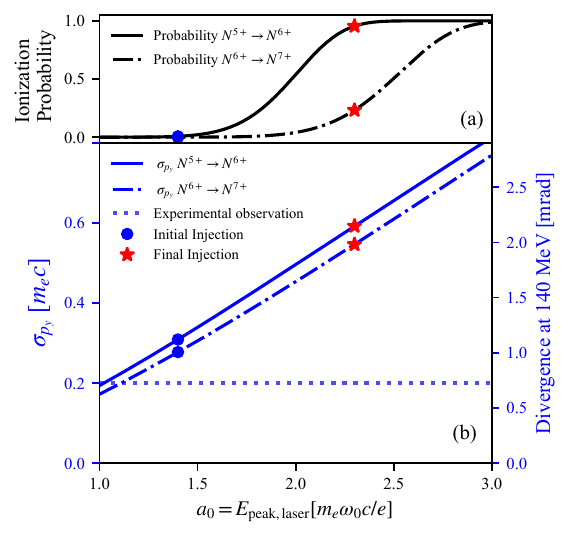} 
		\caption{ (a) ADK tunnel ionization probability in one optical cycle as a function of laser peak normalized field $a_0$; (b) corresponding normalized transverse momentum rms spread $\sigma_{p_y}$ computed from Eq.~\ref{eq:sigmapy}, for both ionization levels. $\sigma_{p_y}$ is shown by a blue line with small dots, as computed from Eq.~\ref{eq:divergence}, assuming $(\theta_y)_{rms}=0.67$ mrad as measured in the experiment.
        The first injection (blue circle) is defined as when the ionization probability exceeds $1\%$ per cycle, the final injection (red star) is defined from the simulated $a_0$ at the point where the nitrogen density used in the PIC simulation is below  $ 2\times 10^{15}$ cm$^{-3}$ (5\% of its peak value)  [see Fig.~\ref{fig:PIC_simulation_bunch_evolution}(a)].} 
		\label{fig:ioni_injection}
	\end{figure}

\begin{figure}[!htbp]
		\centering
			\includegraphics[width=0.99\columnwidth]{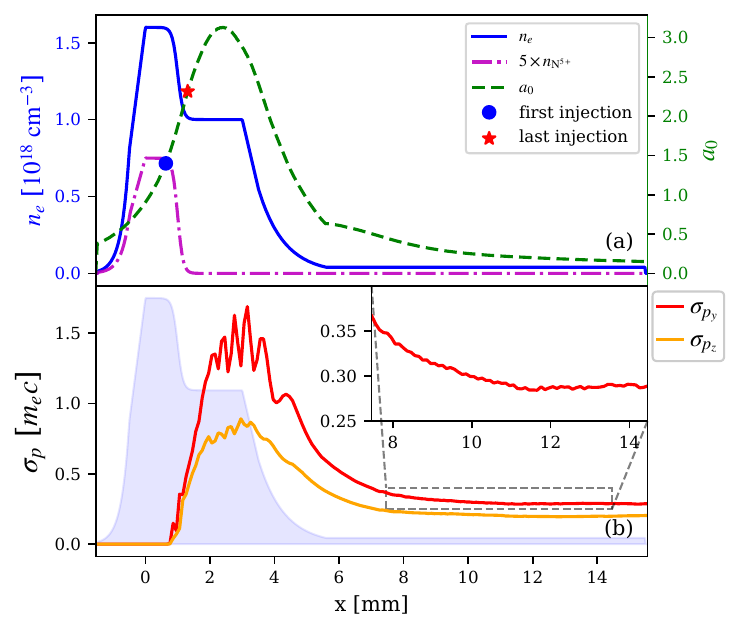} 
		\caption{Simulation results showing evolution of the laser intensity and the transverse momentum spread. (a) electron density profile, for all the electrons (blue) and those from the first five levels of nitrogen (purple); the green line shows the evolution of the laser maximum normalized peak transverse field.   (b) shows the evolution of the momentum spread, perpendicular to the laser polarization direction $\sigma_{p_z}$ (orange) and parallel to polarization $\sigma_{p_y}$ (red).   }
		\label{fig:PIC_simulation_bunch_evolution}
	\end{figure}

The value of the transverse momentum of  accelerated electrons is directly linked to the process of ionization injection. The region of the density profile where injection occurs is limited to the volume where nitrogen is confined and where the laser intensity is above ionization threshold of inner shells. In Fig.~\ref{fig:ioni_injection}(a)  the Ammosov-Delone-Krainov (ADK) \cite{perelomov1966ionization,perelomov1967ionization,ammosov1986tunnel,Chen2013,schroeder_thermal_2014} tunnel ionization probability over one optical cycle for the two innermost shells of nitrogen is plotted as a function of the normalized electric field, $a_0=\mathrm{max}[E_y]/(2\pi m_e c^2/\lambda_0 e)$. Ionization  becomes significant when the probability exceeds 1\%, corresponding here to $a_0 > 1.3$ (blue marker), and we refer to this as the start of ionization. The laser focal plane was chosen in order to have a value of  $a_0$ close to this threshold in the first gas cell compartment, and achieve localized injection and trapping of electrons from these energy levels. This strategy is used to obtain a small initial energy spread and divergence \cite{Jalas2021,Kirchen2021}. Numerical simulations (Fig.~\ref{fig:PIC_simulation_bunch_evolution} (a)) show that $a_0<2.25$ (red marker) during  laser propagation in the region where nitrogen is present.

Inner shell electrons are released near the peaks of the laser transverse electric field, so that
the range of amplitude of the normalized transverse vector potential $a_{0,ioniz}$ (and transverse electric field $E_{0,ioniz}$) at ionization time generates a spread in the transverse momentum of the electrons released through tunnel ionization~\cite{schroeder_thermal_2014}. For a laser pulse linearly polarized along $y$, the value of the transverse momentum rms spread $\sigma_{p_y}$, normalized by $m_ec$, can be quantified as~\cite{schroeder_thermal_2014}:

\begin{equation}
	\sigma_{p_y} = a_{0,ioniz} \Delta  \left[  1 + \left(2n^*-|m|-\frac{11}{2}\right) \Delta^2 \right] ^{1/2} \, , \label{eq:sigmapy}
\end{equation}
where $n^* = Z(U_H/U_I)^{1/2}$, $m$ is the orbital magnetic quantum number, and $\Delta=\left(3E_{0,ioniz}/2E_a\right)^{1/2}\left(U_H/U_I\right)^{3/4}$ as defined in \cite{schroeder_thermal_2014,Massimo2020,Tomassini2022}. Here, $m_e$ is the electron mass, $c$ is the speed of light, $U_I$ and $U_H$ are respectively the ionization potential of the target energy level in the ionized atom and the ionization potential of hydrogen, and $E_a\approx 0.51$ TV/m is the characteristic atomic field. In the following, we  refer to the normalized transverse momentum spread in Eq.~\ref{eq:sigmapy} as the thermal level for the thermal momentum spread in ionization injection. Its value, as a function of $a_{0,ioniz}$, is  shown in Fig.~\ref{fig:ioni_injection}(b), for 
the two nitrogen inner shells. Note that its derivation in \cite{schroeder_thermal_2014} includes all the electrons released through tunnel ionization:  if some of these electrons are not trapped in the laser wakefield, the transverse momentum spread of the trapped electron beam can be smaller. 
The estimated average transverse momentum spread of the measured electron beams  for the ensemble of shots in Fig.~\ref{fig:Waterfalls}(a) is plotted in Fig.~\ref{fig:ioni_injection}(b) as a blue dotted line. 

To get some insight into the physical processes leading to the measured  electron spectra, 
Particle in Cell \cite{BirdsallLangdon2004} (PIC) simulations with the code Smilei \cite{Derouillat2018}, with a laser envelope model\cite{Massimo2019,Massimo2020,Terzani2021} in cylindrical geometry \cite{Lifschitz2009,Massimo2020cylindrical}, were performed laser and plasma parameters based on the experimental values. The longitudinal hydrogen and nitrogen density profiles obtained from the parameterization of the results of the corresponding fluid simulations were used.
 
The laser field distribution was modeled as a Flattened Gaussian Beam (FGB) \cite{Santarsiero1997} with order $N=10$ and with a fundamental mode waist $w_0 = 14 $ \unit{\micro \meter}, and a Gaussian temporal profile with FWHM pulse duration in intensity of $30$~fs. This fit was determined from fluence measurements at different  positions around the focus along the laser axis \cite{Moulanier2023,Moulanier23GSAMD}. With the FGB parameters chosen for the simulation,  a normalized maximum laser field in vacuum of $a_0 = 2.23$ is reached.

\begin{figure}[h]
		\centering
				\includegraphics[width=0.99\columnwidth]{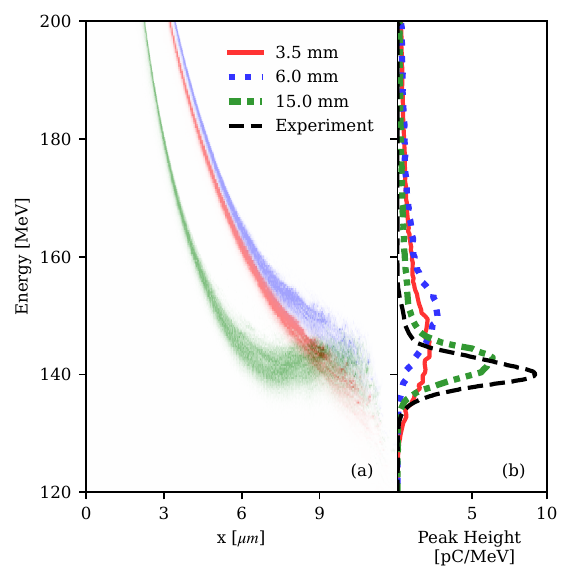} 
			\caption{Evolution of the accelerated electron beam in the plasma exit downramp. (a) Saturated longitudinal phase space distribution in the PIC simulation, at selected positions reported in the legend along density profile. Red: end of the acceleration stage; blue: end of the downramp; green: end of the plasma lens. The longitudinal position is relative to the simulation moving window. (b) corresponding energy spectra, compared with the experimental spectrum shown in Fig.~\ref{fig:Waterfalls}(c) (black dashed line).}
		\label{fig:PIC_simulation_bunch_evolution2}
	\end{figure}

Fig.~\ref{fig:PIC_simulation_bunch_evolution}(a) shows the evolution of the laser normalized peak electric field and the plasma density distribution. The points where the ionization of $N^{5+}$ and $N^{6+}$ starts and ends  as defined in Fig.~\ref{fig:ioni_injection}(a) are also shown. Injection occurs within the first density downramp, which has a length of 500 \unit{\micro\meter}. The decrease of the plasma density in the downramp elongates the plasma wave  and facilitates the trapping of nitrogen electrons from the 6th and 7th energy levels released through ionization \cite{Thaury2015a,Kirchen2021,Dickson2022,Marini2024}.The total charge  $Q_T$ injected into the first bunch (74.4 \unit{\pico\coulomb}) is conserved throughout acceleration in the plateau and after exiting the gas cell. In the LPT,  the electron beam is both transversely focused (Fig.~\ref{fig:PIC_simulation_bunch_evolution}(b)) and spectrally compressed (Fig.~\ref{fig:PIC_simulation_bunch_evolution2}) while the total charge is conserved.

Fig.~\ref{fig:PIC_simulation_bunch_evolution}(b) shows the evolution of the transverse momentum spread in the directions parallel and perpendicular to the laser linear polarization direction $y$. Betatron oscillations are clearly visible in the injection and acceleration regions. For $3<x<6$~mm, the downramp acts as a plasma lens due to the laser driven wakefield, which reduces $\sigma_{p_y} $ to $0.5~m_e c$. For $x>6$~mm, electron beam dynamics are determined by the fact that the beam density exceeds the plasma density, $n_b > n_p$, and the laser intensity has decreased.  The simulation shows that the ratio of the longitudinal electric field driven by the beam exceeds that of the laser by a factor of at least 5 throughout the LPT. However, the laser remains intense enough to ionize the hydrogen gas in front of the beam ($a_0(x=15$~mm$) = 0.2$).  In the region $x >6$mm , $\sigma_{p_y}$ is further reduced to $0.3~m_e c$. 

In Fig.~\ref{fig:PIC_simulation_bunch_evolution2}, the longitudinal phase space distribution (a) and corresponding energy spectra (b) are shown at three longitudinal positions in the LPT. The injection and acceleration processes generate  electron beams with a correlation between longitudinal position and energy, leading to  broad energy spectra. However, in the LPT,  the electron beam drives its own wakefield, which rotates the electron beam distribution in  phase space and leads to a more peaked energy spectrum. The evolution of the energy spectrum started in the exponential downramp (compare spectra at 3.5 and 6~mm in Fig.~\ref{fig:PIC_simulation_bunch_evolution2}(b)) continues smoothly in the LPT after $x=6$~mm and results in a peaked spectrum in good agreement with the one experimentally measured (dotted line).

In conclusion, using a gas cell to tailor a density profile including  a long low density plasma tail at the outlet, we have demonstrated 3D shaping of the laser driven electron beam. For the first time, the transverse momentum spread is shown to be reduced  below the thermal level,  while the energy spread is decreased by rotation in phase space by the beam wakefield in the low density plasma. Using this process,  beams with  40 \unit{\pico\coulomb} (FWHM)  energy spreads of 6 MeV have been achieved. These beams have remarkably low transverse momentum spread of 0.2~$m_ec$, corresponding to divergences of 0.5~\unit{\milli \radian} and consequently maximum spectral brightness of 8 \unit{\pico\coulomb\per{\mega\electronvolt}\per{\milli\radian}}. These results illustrate the potential of using compact, flexible devices to generate high electron beam quality, by reshaping an LWFA electron beam both longitudinally and transversely in the post-acceleration plasma.

\textbf{Acknowledgements}
This project has received funding from the European Union's Horizon 2020 research and innovation programme under grant  agreement no. 871124 Laserlab-Europe. This work was granted access to the HPC
resources of TGCC and CINES under the allocation 2023-A0170510062 (Virtual Laplace) made by GENCI. I.~Moulanier was supported by the CNRS in the framework of the project DIANA, contract N. 1255841. 
The authors would like to thank T. Cloarec for his input in the early stages of this work.

    \bibliographystyle{unsrt} 
	\bibliography{Bibliography.bib}

\providecommand{\noopsort}[1]{}\providecommand{\singleletter}[1]{#1}%
\begin{thebibliography}{10}

\bibitem{Esarey2009}
E.~Esarey, C.~B. Schroeder, and W.~P. Leemans.
\newblock Physics of laser-driven plasma-based electron accelerators.
\newblock {\em Reviews of Modern Physics}, 81(3):1229--1285, 2009.

\bibitem{Hooker2013}
S.~M. Hooker.
\newblock Developments in laser-driven plasma accelerators.
\newblock {\em Nature Photonics}, 7(10):775--782, 2013.

\bibitem{downer_diagnostics_2018}
M.~C. Downer, R.~Zgadzaj, A.~Debus, U.~Schramm, and M.~C. Kaluza.
\newblock Diagnostics for plasma-based electron accelerators.
\newblock {\em Reviews of Modern Physics}, 90(3), August 2018.
\newblock Publisher: American Physical Society (APS).

\bibitem{pak_injection_2010}
A.~Pak, K.~A. Marsh, S.~F. Martins, W.~Lu, W.~B. Mori, and C.~Joshi.
\newblock Injection and {Trapping} of {Tunnel}-{Ionized} {Electrons} into
  {Laser}-{Produced} {Wakes}.
\newblock {\em Physical Review Letters}, 104(2):025003, January 2010.

\bibitem{Pollock2011}
B.~B. Pollock, C.~E. Clayton, J.~E. Ralph, F.~Albert, A.~Davidson, L.~Divol,
  C.~Filip, S.~H. Glenzer, K.~Herpoldt, W.~Lu, K.~A. Marsh, J.~Meinecke, W.~B.
  Mori, A.~Pak, T.~C. Rensink, J.~S. Ross, J.~Shaw, G.~R. Tynan, C.~Joshi, and
  D.~H. Froula.
\newblock Demonstration of a narrow energy spread, $\ensuremath{\sim}0.5\text{
  }\text{ }\mathrm{GeV}$ electron beam from a two-stage laser wakefield
  accelerator.
\newblock {\em Phys. Rev. Lett.}, 107:045001, 2011.

\bibitem{Chen2012}
M.~Chen, E.~Esarey, C.~B. Schroeder, C.~G.~R. Geddes, and W.~P. Leemans.
\newblock Theory of ionization-induced trapping in laser-plasma accelerators.
\newblock {\em Physics of Plasmas}, 19(3):033101, 03 2012.

\bibitem{Mirziae2015}
M.~Mirzaie, S.~Li, M.~Zeng, N.~A.~M. Hafz, M.~Chen, G.~Y. Li, Q.~J. Zhu,
  H.~Liao, T.~Sokollik, F.~Liu, Y.~Y. Ma, L.~M. Chen, Z.~M. Sheng, and
  J.~Zhang.
\newblock Demonstration of self-truncated ionization injection for gev electron
  beams.
\newblock {\em Scientific Reports}, 5(1):14659, 2015.

\bibitem{Thaury2015}
C.~Thaury, E.~Guillaume, A.~Döpp, R.~Lehe, A.~Lifschitz, K.~Ta~Phuoc,
  J.~Gautier, J.~P. Goddet, A.~Tafzi, A.~Flacco, F.~Tissandier, S.~Sebban,
  A.~Rousse, and V.~Malka.
\newblock Demonstration of relativistic electron beam focusing by a
  laser-plasma lens.
\newblock {\em Nature Communications}, 6(1):6860, 2015.

\bibitem{Couperus2017}
J.~P. Couperus, R.~Pausch, A.~K{\"o}hler, O.~Zarini, J.~M. Kr{\"a}mer,
  M.~Garten, A.~Huebl, R.~Gebhardt, U.~Helbig, S.~Bock, K.~Zeil, A.~Debus,
  M.~Bussmann, U.~Schramm, and A.~Irman.
\newblock Demonstration of a beam loaded nanocoulomb-class laser wakefield
  accelerator.
\newblock {\em Nature Communications}, 8(1):487, 2017.

\bibitem{Irman2018}
A.~Irman, J.~P. Couperus, A.~Debus, A.~K{\"o}hler, J.~M. Kr{\"a}mer, R.~Pausch,
  O.~Zarini, and U.~Schramm.
\newblock Improved performance of laser wakefield acceleration by tailored
  self-truncated ionization injection.
\newblock {\em Plasma Physics and Controlled Fusion}, 60(4):044015, 2018.

\bibitem{Kurz2021}
T.~Kurz, T.~Heinemann, M.~F. Gilljohann, Y.~Y. Chang, J.~P. Couperus~Cabadağ,
  A.~Debus, O.~Kononenko, R.~Pausch, S.~Schöbel, R.~W. Assmann, M.~Bussmann,
  H.~Ding, J.~Götzfried, A.~Köhler, G.~Raj, S.~Schindler, K.~Steiniger,
  O.~Zarini, S.~Corde, A.~Döpp, B.~Hidding, S.~Karsch, U.~Schramm, A.~Martinez
  de~la Ossa, and A.~Irman.
\newblock Demonstration of a compact plasma accelerator powered by
  laser-accelerated electron beams.
\newblock {\em Nature Communications}, 12(1):2895, 2021.

\bibitem{Kirchen2021}
Manuel Kirchen, S\"oren Jalas, Philipp Messner, Paul Winkler, Timo Eichner,
  Lars H\"ubner, Thomas H\"ulsenbusch, Laurids Jeppe, Trupen Parikh, Matthias
  Schnepp, and Andreas~R. Maier.
\newblock Optimal beam loading in a laser-plasma accelerator.
\newblock {\em Phys. Rev. Lett.}, 126:174801, 2021.

\bibitem{Chang2023}
Y.-Y. Chang, J.~Couperus Cabada\ifmmode~\breve{g}\else \u{g}\fi{}, A.~Debus,
  A.~Ghaith, M.~LaBerge, R.~Pausch, S.~Sch\"obel, P.~Ufer, U.~Schramm, and
  A.~Irman.
\newblock Reduction of the electron-beam divergence of laser wakefield
  accelerators by integrated plasma lenses.
\newblock {\em Phys. Rev. Appl.}, 20:L061001, 2023.

\bibitem{Picksley2024}
A.~Picksley, J.~Stackhouse, C.~Benedetti, K.~Nakamura, H.~E. Tsai, R.~Li,
  B.~Miao, J.~E. Shrock, E.~Rockafellow, H.~M. Milchberg, C.~B. Schroeder,
  J.~van Tilborg, E.~Esarey, C.~G.~R. Geddes, and A.~J. Gonsalves.
\newblock Matched guiding and controlled injection in dark-current-free,
  10-gev-class, channel-guided laser-plasma accelerators.
\newblock {\em Phys. Rev. Lett.}, 133:255001, 2024.

\bibitem{schroeder_thermal_2014}
C.~B. Schroeder, J.-L. Vay, E.~Esarey, S.~S. Bulanov, C.~Benedetti, L.-L. Yu,
  M.~Chen, C.~G.~R. Geddes, and W.~P. Leemans.
\newblock Thermal emittance from ionization-induced trapping in plasma
  accelerators.
\newblock {\em Physical Review Special Topics - Accelerators and Beams},
  17(10):101301, October 2014.

\bibitem{chen_grand_1987}
Pisin Chen.
\newblock Grand disruption: {A} possible final focussing mechanism for linear
  colliders.
\newblock {\em Particle Accelerators}, 20, 1987.

\bibitem{Kuschel2016}
S.~Kuschel, D.~Hollatz, T.~Heinemann, O.~Karger, M.~B. Schwab, D.~Ullmann,
  A.~Knetsch, A.~Seidel, C.~R\"odel, M.~Yeung, M.~Leier, A.~Blinne, H.~Ding,
  T.~Kurz, D.~J. Corvan, A.~S\"avert, S.~Karsch, M.~C. Kaluza, B.~Hidding, and
  M.~Zepf.
\newblock Demonstration of passive plasma lensing of a laser wakefield
  accelerated electron bunch.
\newblock {\em Phys. Rev. Accel. Beams}, 19:071301, Jul 2016.

\bibitem{Thaury2015lens}
C.~Thaury, E.~Guillaume, A.~D{\"o}pp, R.~Lehe, A.~Lifschitz, K.~Ta~Phuoc,
  J.~Gautier, J-P Goddet, A.~Tafzi, A.~Flacco, F.~Tissandier, S.~Sebban,
  A.~Rousse, and V.~Malka.
\newblock Demonstration of relativistic electron beam focusing by a
  laser-plasma lens.
\newblock {\em Nature Communications}, 6(1):6860, 2015.

\bibitem{Lehe2014}
R.~Lehe, C.~Thaury, E.~Guillaume, A.~Lifschitz, and V.~Malka.
\newblock Laser-plasma lens for laser-wakefield accelerators.
\newblock {\em Phys. Rev. ST Accel. Beams}, 17:121301, Dec 2014.

\bibitem{darcy_tunable_2019}
R.~D’Arcy, S.~Wesch, A.~Aschikhin, S.~Bohlen, C.~Behrens, M.~J. Garland,
  L.~Goldberg, P.~Gonzalez, A.~Knetsch, V.~Libov, A.~Martinez De~La~Ossa,
  M.~Meisel, T.~J. Mehrling, P.~Niknejadi, K.~Poder, J.-H. Röckemann,
  L.~Schaper, B.~Schmidt, S.~Schröder, C.~Palmer, J.-P. Schwinkendorf,
  B.~Sheeran, M.~J.~V. Streeter, G.~Tauscher, V.~Wacker, and J.~Osterhoff.
\newblock Tunable {Plasma}-{Based} {Energy} {Dechirper}.
\newblock {\em Physical Review Letters}, 122(3):034801, January 2019.

\bibitem{wu_phase_2019}
Y.~P. Wu, J.~F. Hua, Z.~Zhou, J.~Zhang, S.~Liu, B.~Peng, Y.~Fang, Z.~Nie, X.~N.
  Ning, C.-H. Pai, Y.~C. Du, W.~Lu, C.~J. Zhang, W.~B. Mori, and C.~Joshi.
\newblock Phase {Space} {Dynamics} of a {Plasma} {Wakefield} {Dechirper} for
  {Energy} {Spread} {Reduction}.
\newblock {\em Physical Review Letters}, 122(20):204804, May 2019.

\bibitem{drobniak_two-chamber_2023}
P.~Drobniak, E.~Baynard, K.~Cassou, D.~Douillet, J.~Demailly, A.~Gonnin,
  G.~Iaquaniello, G.~Kane, S.~Kazamias, N.~Lericheux, B.~Lucas, B.~Mercier,
  Y.~Peinaud, and M.~Pittman.
\newblock Two-chamber gas target for laser-plasma accelerator electron source,
  September 2023.
\newblock arXiv:2309.11921 [physics].

\bibitem{weller1998}
HG~Weller, G~Tabor, H~Jasak, and C~Fureby.
\newblock A tensorial approach to computational continuum mechanics using
  object-oriented techniques.
\newblock {\em Computers in Physics}, 12(6):620--631, 1998.

\bibitem{openfoam}
{OpenFOAM Foundation}.
\newblock {OpenFOAM: The Open Source CFD Toolbox}, 2025.

\bibitem{perelomov1966ionization}
A.~M. Perelomov, V.~S. Popov, and M.~V. Terent'ev.
\newblock Ionization of atoms in an alternating electric field.
\newblock {\em Soviet Physics JETP}, 23:924--934, 1966.

\bibitem{perelomov1967ionization}
A.~M. Perelomov, V.~S. Popov, and M.~V. Terent'ev.
\newblock Ionization of atoms in an alternating electric field ii.
\newblock {\em Soviet Physics JETP}, 24:207--217, 1967.

\bibitem{ammosov1986tunnel}
M.~V. Ammosov, N.~B. Delone, and V.~P. Krainov.
\newblock Tunnel ionization of complex atoms and of atomic ions in an
  alternating electromagnetic field.
\newblock {\em Soviet Physics JETP}, 64:1191--1194, 1986.

\bibitem{Chen2013}
M.~Chen, E.~Cormier-Michel, C.G.R. Geddes, D.L. Bruhwiler, L.L. Yu, E.~Esarey,
  C.B. Schroeder, and W.P. Leemans.
\newblock Numerical modeling of laser tunneling ionization in explicit
  particle-in-cell codes.
\newblock {\em Journal of Computational Physics}, 236:220--228, 2013.

\bibitem{Jalas2021}
S{\"o}ren Jalas, Manuel Kirchen, Philipp Messner, Paul Winkler, Lars
  H{\"u}bner, Julian Dirkwinkel, Matthias Schnepp, Remi Lehe, and Andreas~R
  Maier.
\newblock Bayesian optimization of a laser-plasma accelerator.
\newblock {\em Phys. Rev. Lett.}, 126(10):104801, 2021.

\bibitem{Massimo2020}
F.~Massimo, A.~Beck, J.~Derouillat, I.~Zemzemi, and A.~Specka.
\newblock Numerical modeling of laser tunneling ionization in particle-in-cell
  codes with a laser envelope model.
\newblock {\em Phys. Rev. E}, 102:033204, Sep 2020.

\bibitem{Tomassini2022}
Paolo Tomassini, Francesco Massimo, Luca Labate, and Leonida~A. Gizzi.
\newblock Accurate electron beam phase-space theory for ionization-injection
  schemes driven by laser pulses.
\newblock {\em High Power Laser Science and Engineering}, 10:e15, 2022.

\bibitem{BirdsallLangdon2004}
C.~K. Birdsall and A.~B. Langdon.
\newblock {\em Plasma Physics via Computer Simulation}.
\newblock Taylor and Francis Group, 2004.

\bibitem{Derouillat2018}
J.~Derouillat, A.~Beck, F.~Pérez, T.~Vinci, M.~Chiaramello, A.~Grassi,
  M.~Flé, G.~Bouchard, I.~Plotnikov, N.~Aunai, J.~Dargent, C.~Riconda, and
  M.~Grech.
\newblock Smilei : A collaborative, open-source, multi-purpose particle-in-cell
  code for plasma simulation.
\newblock {\em Computer Physics Communications}, 222:351--373, 2018.

\bibitem{Massimo2019}
F~Massimo, A~Beck, J~Derouillat, M~Grech, M~Lobet, F~P{\'e}rez, I~Zemzemi, and
  A~Specka.
\newblock Efficient start-to-end 3d envelope modeling for two-stage laser
  wakefield acceleration experiments.
\newblock {\em Plasma Physics and Controlled Fusion}, 61(12):124001, oct 2019.

\bibitem{Terzani2021}
D.~Terzani, C.~Benedetti, C.~B. Schroeder, and E.~Esarey.
\newblock Accuracy of the time-averaged ponderomotive approximation for
  laser-plasma accelerator modeling.
\newblock {\em Physics of Plasmas}, 28(6):063105, 06 2021.

\bibitem{Lifschitz2009}
A.F. Lifschitz, X.~Davoine, E.~Lefebvre, J.~Faure, C.~Rechatin, and V.~Malka.
\newblock Particle-in-cell modelling of laser--plasma interaction using fourier
  decomposition.
\newblock {\em Journal of Computational Physics}, 228(5):1803--1814, 2009.

\bibitem{Massimo2020cylindrical}
F.~Massimo, I.~Zemzemi, A.~Beck, J.~Derouillat, and A.~Specka.
\newblock Efficient cylindrical envelope modeling for laser wakefield
  acceleration.
\newblock {\em Journal of Physics: Conference Series}, 1596(1):012055, jul
  2020.

\bibitem{Santarsiero1997}
M.~Santarsiero, D.~Aiello, R.~Borghi, and S.~Vicalvi and.
\newblock Focusing of axially symmetric flattened gaussian beams.
\newblock {\em Journal of Modern Optics}, 44(3):633--650, 1997.

\bibitem{Moulanier2023}
I.~Moulanier, L.~T. Dickson, C.~Ballage, O.~Vasilovici, A.~Gremaud,
  S.~Dobosz~Dufrénoy, N.~Delerue, L.~Bernardi, A.~Mahjoub, A.~Cauchois,
  A.~Specka, F.~Massimo, G.~Maynard, and B.~Cros.
\newblock {Modeling of the driver transverse profile for laser wakefield
  electron acceleration at APOLLON research facility}.
\newblock {\em Physics of Plasmas}, 30(5):053109, 2023.

\bibitem{Moulanier23GSAMD}
I.~Moulanier, L.~T. Dickson, F.~Massimo, G.~Maynard, and B.~Cros.
\newblock Fast laser field reconstruction method based on a
  gerchberg\&\#x2013;saxton algorithm with mode decomposition.
\newblock {\em J. Opt. Soc. Am. B}, 40(9):2450--2461, 2023.

\bibitem{Thaury2015a}
C.~Thaury, E.~Guillaume, A.~Lifschitz, K.~Ta~Phuoc, M.~Hansson, G.~Grittani,
  J.~Gautier, J.~P. Goddet, A.~Tafzi, O.~Lundh, and V.~Malka.
\newblock Shock assisted ionization injection in laser-plasma accelerators.
\newblock {\em Scientific Reports}, 5(1):16310, 2015.

\bibitem{Dickson2022}
LT~Dickson, CID Underwood, F~Filippi, RJ~Shalloo, J~Bj{\"o}rklund Svensson,
  D~Gu{\'e}not, K~Svendsen, I~Moulanier, S~Dobosz Dufr{\'e}noy, CD~Murphy,
  et~al.
\newblock Mechanisms to control laser-plasma coupling in laser wakefield
  electron acceleration.
\newblock {\em Physical Review Accelerators and Beams}, 25(10):101301, 2022.

\bibitem{Marini2024}
Samuel Marini, Damien F.~G. Minenna, Francesco Massimo, Laury Batista, Vittorio
  Bencini, Antoine Chanc\'e, Nicolas Chauvin, Steffen Doebert, John Farmer,
  Edda Gschwendtner, Ioaquin Moulanier, Patric Muggli, Didier Uriot, Brigitte
  Cros, and Phu Anh~Phi Nghiem.
\newblock Beam physics studies for a high charge and high beam quality
  laser-plasma accelerator.
\newblock {\em Phys. Rev. Accel. Beams}, 27:063401, Jun 2024.

\end{thebibliography}


\providecommand{\noopsort}[1]{}\providecommand{\singleletter}[1]{#1}%
\begin{thebibliography}{10}

\bibitem{BirdsallLangdon2004}
C.~K. Birdsall and A.~B. Langdon.
\newblock {\em Plasma Physics via Computer Simulation}.
\newblock Taylor and Francis Group, 2004.

\bibitem{Derouillat2018}
J.~Derouillat, A.~Beck, F.~Pérez, T.~Vinci, M.~Chiaramello, A.~Grassi,
  M.~Flé, G.~Bouchard, I.~Plotnikov, N.~Aunai, J.~Dargent, C.~Riconda, and
  M.~Grech.
\newblock Smilei : A collaborative, open-source, multi-purpose particle-in-cell
  code for plasma simulation.
\newblock {\em Computer Physics Communications}, 222:351--373, 2018.

\bibitem{Terzani2020}
Davide Terzani, Pasquale Londrillo, Paolo Tomassini, and Leonida~A Gizzi.
\newblock Numerical implementation of a hybrid pic-fluid framework in
  laser-envelope approximation.
\newblock {\em Journal of Physics: Conference Series}, 1596(1):012062, jul
  2020.

\bibitem{Terzani2021}
D.~Terzani, C.~Benedetti, C.~B. Schroeder, and E.~Esarey.
\newblock Accuracy of the time-averaged ponderomotive approximation for
  laser-plasma accelerator modeling.
\newblock {\em Physics of Plasmas}, 28(6):063105, 2021.

\bibitem{Massimo2019}
F~Massimo, A~Beck, J~Derouillat, M~Grech, M~Lobet, F~P{\'{e}}rez, I~Zemzemi,
  and A~Specka.
\newblock Efficient start-to-end 3d envelope modeling for two-stage laser
  wakefield acceleration experiments.
\newblock {\em Plasma Physics and Controlled Fusion}, 61(12):124001, oct 2019.

\bibitem{Massimo2019cylindrical}
Francesco Massimo, Imen Zemzemi, Arnaud Beck, Julien Dérouillat, and Arnd
  Specka.
\newblock Efficient cylindrical envelope modeling for laser wakefield
  acceleration, 2019.

\bibitem{Massimo2020}
F.~Massimo, A.~Beck, J.~Derouillat, I.~Zemzemi, and A.~Specka.
\newblock Numerical modeling of laser tunneling ionization in particle-in-cell
  codes with a laser envelope model.
\newblock {\em Phys. Rev. E}, 102:033204, Sep 2020.

\bibitem{Lifschitz2009}
A.~Lifschitz, X.~Davoine, E.~Lefebvre, J.~Faure, C.~Rechatin, and V.~Malka.
\newblock {Particle-in-Cell modelling of laser--plasma interaction using
  Fourier decomposition}.
\newblock {\em {Journal of Computational Physics}}, 228(5):1803--1814, November
  2008.

\bibitem{Santarsiero1997}
M.~Santarsiero, D.~Aiello, R.~Borghi, and S.~Vicalvi and.
\newblock Focusing of axially symmetric flattened gaussian beams.
\newblock {\em Journal of Modern Optics}, 44(3):633--650, 1997.

\bibitem{weller1998}
HG~Weller, G~Tabor, H~Jasak, and C~Fureby.
\newblock A tensorial approach to computational continuum mechanics using
  object-oriented techniques.
\newblock {\em Computers in Physics}, 12(6):620--631, 1998.

\bibitem{openfoam}
{OpenFOAM Foundation}.
\newblock {OpenFOAM: The Open Source CFD Toolbox}, 2025.

\bibitem{boris:proc1973}
J.P. Boris.
\newblock Relativistic plasma simulation - optimization of a hybrid code.
\newblock {\em Proceedings, Fourth Conference on the Numerical Simulation of
  Plasma}, 1970.

\bibitem{ADK1986}
Maxim~V Ammosov, Nikolai~B Delone, and Vladimir~P Krainov.
\newblock {Tunnel Ionization Of Complex Atoms And Atomic Ions In
  Electromagnetic Field}.
\newblock In John~A. Alcock, editor, {\em High Intensity Laser Processes},
  volume 0664, pages 138 -- 141. International Society for Optics and
  Photonics, SPIE, 1986.

\bibitem{esirkepov:CPC2001}
T.Zh. Esirkepov.
\newblock Exact charge conservation scheme for particle-in-cell simulation with
  an arbitrary form-factor.
\newblock {\em Computer Physics Communications}, 135(2):144 -- 153, 2001.

\bibitem{Yee1966}
K.~{Yee}.
\newblock {Numerical solution of inital boundary value problems involving
  maxwell's equations in isotropic media}.
\newblock {\em IEEE Transactions on Antennas and Propagation}, 14:302--307, May
  1966.

\bibitem{Terzani2019}
Davide Terzani and Pasquale Londrillo.
\newblock A fast and accurate numerical implementation of the envelope model
  for laser--plasma dynamics.
\newblock {\em Computer Physics Communications}, 242:49 -- 59, 2019.

\bibitem{Bourgeois2023}
Pierre-Louis Bourgeois and Xavier Davoine.
\newblock Improved modellisation of laser–particle interaction in
  particle-in-cell simulations.
\newblock {\em Journal of Plasma Physics}, 89(2):905890206, 2023.

\bibitem{Bouchard2024PML}
Guillaume Bouchard, Arnaud Beck, Francesco Massimo, and Arnd Specka.
\newblock {Perfectly Matched Layer implementation for E-H fields and Complex
  Wave Envelope propagation in the Smilei PIC code}, 2024.
\newblock Submitted to Computer Physics Communications.

\end{thebibliography}

\end{document}


\title{Observation of  laser plasma accelerated electrons with  transverse momentum spread below the thermal level}

\author{T.L. Steyn$^1$ \and A. Panchal$^2$ \and O. Vasilovici$^1$ \and S. Sch\"obel$^3$ \and P. Ufer$^3$ \and F.M. Herrmann$^3$ \and Y.-Y. Chang$^3$ \and I. Moulanier$^1$ \and M. Masckala$^1$ \and O. Khomyshyn$^1$ \and C. Ballage$^1$ \and M. LaBerge$^3$ \and F. Massimo$^1$ \and S. Dobosz Dufrénoy$^2$ \and U. Schramm$^3$ \and A. Irman$^3$ \and B. Cros$^1$}
\maketitle

$^1$ LPGP, CNRS Université Paris Saclay, 91400 Orsay \\ France  $^2$ Université Paris-Saclay, CEA, LIDYL, 91191 Gif sur Yvette, France \\ %
{$^3$ Helmholtz-Zentrum Dresden Rossendorf, Bautzner Landstraße 400, 01328 Dresden, Germany }

theunis-lodewyk.steyn@universite-paris-saclay. \\

\subsection{Supplemental Material Relating to Simulation}

The settings of the Particle in Cell (PIC) \cite{BirdsallLangdon2004} simulation shown in Fig. 4,5 of the article are detailed in the following. 

The simulation was performed with the open source PIC code Smilei (\cite{Derouillat2018}), version 5.1. 

The laser and plasma parameters were varied within the experimental uncertainties in order to recreate the mechanisms shown in the paper. This parameter space exploration was performed using a quick laser envelope model \cite{Terzani2020,Terzani2021,Massimo2019,Massimo2019cylindrical,Massimo2020}, but the simulation shown in the article does not use this approximation.

The simulation was performed in quasi-cylindrical geometry with azimuthal Fourier decomposition \cite{Lifschitz2009}, using two azimuthal modes. To follow the propagation of the laser pulse, a simulation window moving along the $x$ direction at speed $c$ was used. The grid cell size is $\Delta x=$ 0.019~$\mu$m and $\Delta r=$ 0.35$\mu$m in the longitudinal and radial direction respectively.
The integration timestep was set to $c\Delta t/\Delta x=$ $0.975\times dx /c = 0.06327 $  and the simulation window size was $n_x= 3584$ and $n_r=512$ grid cells longitudinal and radial direction respectively 

The laser pulse, linearly polarized in the $y$ direction, was modeled with a Flattened Gaussian beam \cite{Santarsiero1997} in the transverse direction and a Gaussian temporal profile. The order $N=10$, fundamental mode waist $w_0=$ $ 14 \mu$m and $a_0= 2.27$ and Full Width Half Maximum duration $T=30$ fs were chosen to fit the experimental measurements.

Due to the relatively low ionization potential of hydrogen and of the first levels of nitrogen, the plasma was assumed to be composed by already ionized hydrogen and nitrogen ionized up to the 5th level. The free electrons and the ions were modeled with $[1,1,8]$ macro-particles per cell distributed regularly in the $x$, $r$ directions and along the azimuthal angle interval $\theta=[0,2\pi)$ respectively. The plasma was assumed as initially uniform in the radial direction and with a density profile shown in the article Figs 1 and 4, that was obtained from the parametrization of the density distribution obtained with OpenFOAM simulations \cite{weller1998,openfoam} of the gas cell filling. The macroparticles are advanced in the phase space using the Boris scheme \cite{boris:proc1973}. The $N^{5+}$ ion macro-particles, assumed immobile, were subject to tunnel ionization using the Ammosov, Delone, and Krainov (ADK) ionization rate \cite{ADK1986,Massimo2020}. 

The simulation used some numerical schemes in order to increase its accuracy, namely:
\begin{itemize}
\item a charge conserving current deposition scheme \cite{esirkepov:CPC2001}.
\item a Yee-like \cite{Yee1966} solver for Maxwell's equations with low numerical dispersion \cite{Terzani2019}, adapted to the cylindrical geometry. 
\item a B-TIS3 interpolation scheme \cite{Bourgeois2023} to reduce the effects of numerical Cherenkov radiation and the numerical artifacts due to the space-time staggering of the electromagnetic fields in Yee-like Maxwell solvers, adapted to the cylindrical geometry. 
\item a perfectly matched layer boundary condition (number of pml cells            = [[0,0],[30,30]],) on the transverse simulation window border \cite{Bouchard2024PML}.

\end{itemize}

This work was granted access
to the HPC resources of TGCC and CINES under the
allocation 2023-A0170510062 (Virtual Laplace) made by
GENCI

\bibliographystyle{unsrt} 
\bibliography{Bibliography_Supp}